# Overcoming $Si_3N_4$ film stress limitations for High Quality factor ring resonators

Kevin Luke[1], Avik Dutt[1], Carl B. Poitras[1], and Michal Lipson[1,*]

[1]*School of Electrical and Computer Engineering, Cornell University, Ithaca, NY 14853, USA*
[*]*ml292@cornell.edu*

**Abstract:** Silicon nitride ($Si_3N_4$) ring resonators are critical for a variety of photonic devices. However the intrinsically high film stress of silicon nitride has limited both the optical confinement and quality factor (Q) of ring resonators. We show that stress in $Si_3N_4$ films can be overcome by introducing mechanical trenches for isolating photonic devices from propagating cracks. We demonstrate a $Si_3N_4$ ring resonator with an intrinsic quality factor of 7 million, corresponding to a propagation loss of 4.2 dB/m. This is the highest quality factor reported to date for high confinement $Si_3N_4$ ring resonators in the 1550 nm wavelength range.

Silicon nitride ($Si_3N_4$) ring resonators are critical for efficient and compact on chip optical routing [1–3], frequency combs [4–7], and high precision sensing [8–11], however the intrinsically high film stress of silicon nitride has limited both the optical confinement and quality factor (Q) of ring resonators. Whereas the silicon and silicon dioxide platforms generally suffer from high losses or delocalized optical modes, the $Si_3N_4$ platform provides advantages of both high confinement and high Q. High Q disks have also been demonstrated in $Si_3N_4$ [12] but disks have larger mode volumes and are challenging to dispersion engineer for nonlinear applications. $Si_3N_4$ is also a deposited material, which enables seamless integration with other material platforms. However, the high film stress of $Si_3N_4$ prevents thick (>400 nm) films from being deposited; catastrophic cracking occurs, severely limiting device yield.

Thick films, limited to date by stress, would enable high confinement and high Q. Thick films lead to smaller optical mode overlap with the boundaries of the waveguides (responsible for scattering losses) leading to lower losses. This can be observed in Fig. 1 where we show the mode profile for a mode confined in two different waveguide thicknesses. Fig. 1a shows the mode confined in a waveguide with a traditional thickness of 400 nm limited to date by stress. One can see that the mode overlaps significantly with the boundaries of the waveguide. Fig. 1b shows the mode confined in a waveguide with a much higher thickness of 910 nm. One can see that very little of the mode overlaps with the boundary of the waveguide.

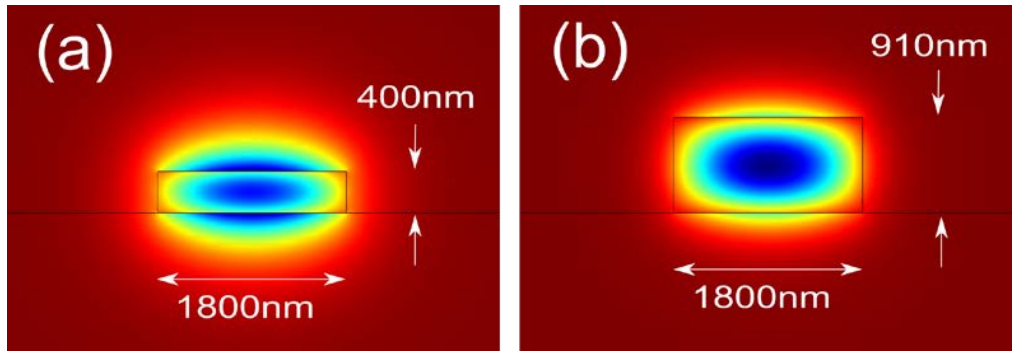

Fig. 1. Transverse magnetic (TM) mode simulations at 1550 nm wavelength for the (a) 400 nm x 1800 nm and (b) 910 nm x 1800 nm waveguides with 55% and 93% modal confinement, respectively. Less of the optical field interacts with the waveguide boundaries for the taller waveguide (b).

Previous high Q $Si_3N_4$ ring resonators have circumvented film stress issues by exploiting either highly delocalized optical modes with extremely thin films [13] or highly confined modes within film stress limits [14]. High Q ring resonators based on extremely thin $Si_3N_4$ films can avoid film stress issues, but they suffer from highly delocalized optical modes, requiring millimeter-scale bending radii and up to 15 µm of silicon oxide cladding. In addition, these resonators only support the transverse electric (TE) mode which prevents integration with devices that support the transverse magnetic (TM) mode. High confinement ring resonators based on thicker films can be achieved using high temperature deposition and anneal to relieve film stress, but films greater than 750 nm in thickness remain challenging.

In order to overcome the stress limitations of $Si_3N_4$, we strategically place mechanical trenches to isolate the photonic devices from propagating cracks. Physical shocks near the edge of the wafer, which occur often during handling of the wafer, can provide enough energy to induce cracking of the stressed film. Once initiated, these cracks originating from the edge of the wafer propagate continuously in a uniform stress field, terminating only once they encounter crack resistance at the edge of the wafer or at another crack boundary. We introduce trenches around our devices that terminate cracks before they can spread to our device region (Fig. 2). A single trench does not guarantee crack termination however [15].

Overstressed films store energy in the enhanced acceleration of the crack and the extended penetration into the substrate. With this stored energy, cracks can overcome the crack resistance of a single trench and continue propagation. To increase crack resistance, we create between two and five parallel trenches to ensure crack termination.

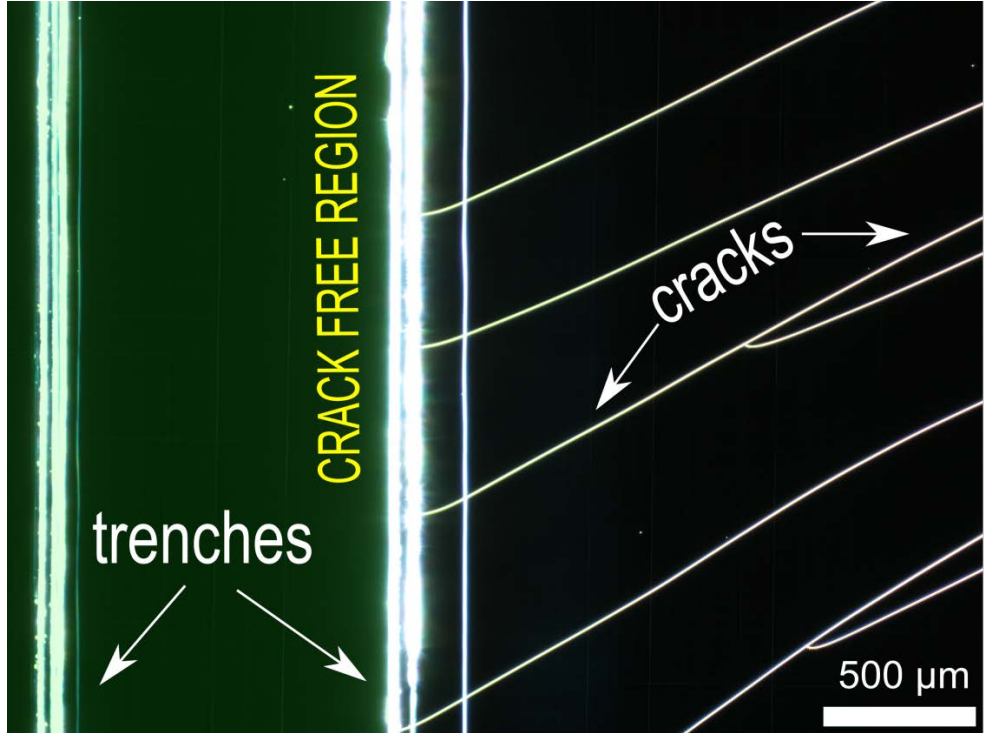

Fig. 2. Microscope images in the dark field showing crack propagation terminating at a trench created with a diamond scribe. The film to the left of these trenches is crack free.

We define a series of trenches around the edge of the wafer before deposition of the $Si_3N_4$ film to prevent crack propagation. Beginning with a thermally oxidized silicon wafer, we lightly scribe a series of lines into the silicon oxide surface to define a 5 cm by 5 cm rectangular region in the center of the wafer. Note that for ease of processing and throughput we define trenches with a diamond scribe, but in principle one can define more sophisticated trenches with photolithography and etching. Etching may actually improve the crack resistance of these trenches because of the increased roughness on the etched surfaces [15]. Following trench definition, we proceed with device fabrication as described in [14]. After deposition of 910 nm of $Si_3N_4$, in steps of 400 nm and 510 nm, we pattern devices with electron beam lithography using ma-N 2405 resist, post exposure bake for 5 minutes at 115°C, and etch in an inductively coupled plasma reactive ion etcher (ICP RIE) using $CHF_3/O_2$ chemistry. After stripping the resist, we anneal devices at 1200°C in a nitrogen atmosphere for 3 hours. We clad devices with 250 nm of high temperature silicon dioxide (HTO) deposited at 800°C followed by 2 µm deposition of silicon dioxide using plasma enhanced chemical vapor deposition (PECVD).

We measure an intrinsic quality factor of 7 million, the highest quality factor reported to date for high confinement $Si_3N_4$ ring resonators; this quality factor corresponds to an ultra-low propagation loss of 4.2 dB/m. We couple a tunable laser light source, transmitted through a polarization controller, into the inverse nanotaper of our device using a lensed fiber. We collect the output of the ring resonator through another inverse nanotaper and collimating lens. After passing the output through a polarizer, we monitor the output on a photodetector. In order to measure a single resonance, we finely scan the laser frequency by applying a triangular-wave voltage signal to the piezoelectric transducer of the laser, while monitoring the photodetector signal on an oscilloscope. We calibrate the voltage-frequency conversion with respect to a free space bowtie cavity. From the transmission spectrum, we measure a resonance linewidth of 36 MHz and resonant transmission of 26% (Fig. 3), corresponding to an intrinsic quality factor of 7 million for TM polarization. For TE polarization we measure a Q of 4 million. We also calculate the propagation loss within the ring given the relation [16]

$$\alpha = \frac{2\pi n_g}{Q \lambda_0} = \frac{\lambda_0}{Q \cdot R \cdot FSR}, \quad (1)$$

where $n_g$ is the group index, $\lambda_0$ is the resonant wavelength, $R$ is the radius of the ring resonator, and *FSR* is the free spectral range. For our device with ring radius of 115 µm, *FSR* of 2.0 nm, and resonant wavelength of 1554.8 nm, we calculate a propagation loss of 4.2 dB/m. This is the lowest loss reported for high confinement waveguides.

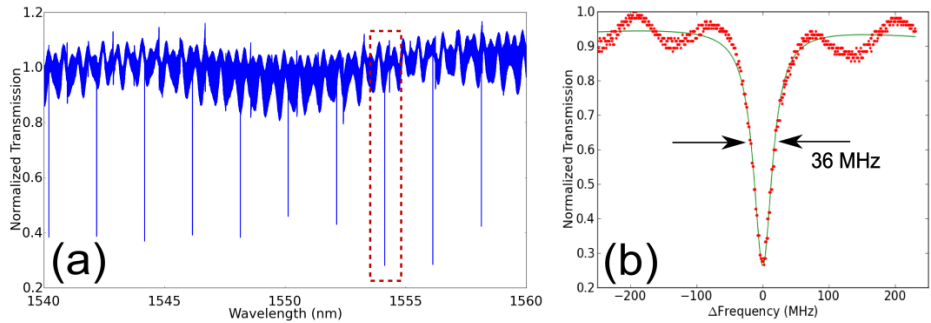

Fig. 3. (a) Transmission spectrum with the finely scanned resonance outlined in red. (b) Resonance with 36 MHz linewidth corresponding to an intrinsic Q of 7 million.

We demonstrate a high quality factor of 7 million in a high confinement $Si_3N_4$ ring resonator using crack resistant trenches to overcome stress limitations of thick $Si_3N_4$ films. Our high Q devices herald advances in low loss optical routing, low power threshold nonlinear optics, and high sensitivity sensors. We have also overcome film stress limitations for $Si_3N_4$, revealing a new design space for integrated optics and microelectromechanical (MEMS) devices that has been unexplored to date.

**Acknowledgements**

This work was supported in part by the Cornell Center for Materials Research with funding from A Graduate Traineeship in Materials for a Sustainable Future, (DGE-0903653). The authors gratefully acknowledge support from DARPA for award # W911NF-11-1-0202 supervised by Dr. Jamil Abo-Shaeer, AFOSR for award # BAA-AFOSR-2012-02 supervised by Dr. Enrique Parra, and the Defense Advanced Research Projects Agency (DARPA) under award #FA8650-10-1-7064.